\documentclass[aoas]{imsart}

\RequirePackage[OT1]{fontenc}
\RequirePackage{amsthm,amsmath,amsfonts}
\RequirePackage{natbib}
\RequirePackage[colorlinks,citecolor=blue,urlcolor=blue]{hyperref}
\usepackage{tikz}

\usetikzlibrary{arrows,chains,matrix,positioning,scopes}
\makeatletter
\tikzset{join/.code=\tikzset{after node path={%
			\ifx\tikzchainprevious\pgfutil@empty\else(\tikzchainprevious)%
			edge[every join]#1(\tikzchaincurrent)\fi}}}
\makeatother
\tikzset{>=stealth',every on chain/.append style={join},
	every join/.style={->}}
\tikzstyle{labeled}=[execute at begin node=$\scriptstyle,
execute at end node=$]

\startlocaldefs
\numberwithin{equation}{section}
\theoremstyle{plain}

\endlocaldefs

\begin{document}
	
	\begin{frontmatter}
		\title{Modeling and Estimation for Self-Exciting Spatio-Temporal Models of Terrorist Activity}
		\runtitle{Self-Exciting Spatio-Temporal Models}

		\begin{aug}
			\author{\fnms{Nicholas J.} \snm{Clark\thanksref{t1}}\ead[label=e1]{nclark1@iastate.edu}} \and
			\author{\fnms{Philip M.} \snm{Dixon}\ead[label=e2]{pdixon@iastate.edu}}

			\thankstext{t1}{Supported in Part through a Omar Nelson Bradley Fellowship}
		
			\runauthor{N. Clark and P. Dixon}
			
			\affiliation{Iowa State University} 
			
			\address{Iowa State University\\
				Department of Statistics\\
				\printead{e1}\\
				\phantom{E-mail:\ nclark1@iastate.edu}\\
				\printead*{e2}}

		\end{aug}
		
		\begin{abstract}
			Spatio-temporal hierarchical modeling is an extremely attractive way to model the spread of crime or terrorism data over a given region, especially when the observations are counts and must be modeled discretely.  The spatio-temporal diffusion is placed, as a matter of convenience, in the process model allowing for straightforward estimation of the diffusion parameters through Bayesian techniques.  However, this method of modeling does not allow for the existence of self-excitation, or a temporal data model dependency, that has been shown to exist in criminal and terrorism data.  In this manuscript we will use existing theories on how violence spreads to create models that allow for both spatio-temporal diffusion in the process model as well as temporal diffusion, or self-excitation, in the data model.  We will further demonstrate how Laplace approximations similar to their use in Integrated Nested Laplace Approximation can be used to quickly and accurately conduct inference of self-exciting spatio-temporal models allowing practitioners a new way of fitting and comparing multiple process models.  We will illustrate this approach by fitting a self-exciting spatio-temporal model to terrorism data in Iraq and demonstrate how choice of process model leads to differing conclusions on the existence of self-excitation in the data and differing conclusions on how violence spread spatially-temporally in that country from 2003-2010.
			
		\end{abstract}
		
	
		
	\end{frontmatter}
	
	
	\section{Introduction} 
	A typical spatio-temporal model consists of three levels, a data model, a process model, and a parameter model.  A common way to model data then is to assume $Y(\cdot)$, is conditionally independent given the process model $X(\cdot)$. For example, if observations take place at aerial regions, $\boldsymbol{s_i}$, at discrete time periods, $t$, and $Y(\boldsymbol{s_i},t)$ are counts, a common model is $Y(\boldsymbol{s_i},t)|X(\boldsymbol{s_i},t)\sim \mbox{Pois}(\exp(X(\boldsymbol{s_i},t)))$.  The spatio-temporal diffusion structure is commonly then placed on the process model which commonly is assumed to have a Gaussian joint distribution of $\boldsymbol{X}\sim \mbox{Gaus}(\boldsymbol{0},Q^{-1}(\theta))$.  The majority of analysis of these models is done using Bayesian techniques requiring a further parameter model for $\theta$.  The challenge in these models is, then, determining an appropriate structure for $\boldsymbol{Q}^{-1}(\theta)$ or $\boldsymbol{Q}(\theta)$.  If both the covariance and the precision is chosen to be too dense inference quickly becomes impossible due to the size of $\boldsymbol{Q}^{-1}(\theta)$.  In spatio-temporal models it is quite common for the dimension of $\boldsymbol{Q}$ to be larger than $10^4\times 10^4$.  A thorough overview giving many examples of this method of modeling is given in \cite{cressie2015statistics}.
	
	In modeling terrorism or crime data one possibility is to use an extremely general spatio-temporal process model to capture variance not explained through the use of covariates.  For example \cite{python2016bayesian} use a Matern class covariance function over space and an AR(1) process over time.  They then use covariates to test the impact of infrastructure, population, and governance.  The general spatio-temporal process models used, in this case, has an extremely sparse precision structure greatly aiding in computations.
	
	While diffusion in spatio-temporal models is often modeled through a latent process, more recent models describing the spread of violence have incorporated self-excitation, or spatio-temporal diffusion that exists linearly in the data model itself.  Self-excitation is the theory that in terrorism, or crime, the probability of an event occurring is a function of previous successful events.  For instance \cite{mohler2012self} demonstrate that burglars are more likely to rob locations that have previously, successfully, been robbed.  \cite{mohler2013modeling} derived a class of models that allowed for temporal diffusion in both the process model as well as the data model and demonstrated how the two processes were identifiable.
	
	In the modeling of terrorism data \cite{lewis2012self}, \cite{porter2012self}, and \cite{mohler2013modeling} have all successfully used the self-excitation approach to model.  Most recently, \cite{tench2016spatio} used a likelihood approach for temporal modeling of IEDs in Northern Ireland using self-excitation.  However, in these papers, the existence and analysis of self-excitation was the primary objective and any process model dependency was either ignored or treated as a nuisance.  The one exception is in \cite{mohler2013modeling} where a temporal only model was assumed for the process model and inference was conducted on both the process model dependency and the data model dependency.
	
	In this manuscript, we will consider a spatial and a spatial-temporal process model that allows for self-excitation.  We will present two self-exciting models for terrorist activity that have different process models corresponding to different notions of how terrorism evolves in time and space as well as temporal dependency in the data model to account for self-excitation.  These two models are specific cases of more general spatio-temporal models that allow dependency in both the process model as well as the data model.
	
	We will further show how Laplace approximations similar to their use in Integrated Nested Laplace Approximation, or INLA, an approximate Bayesian method due to \cite{rue2009approximate} can be used to conduct inference for these types of models. We will show, via simulation, how INLA, when appropriately modified, can accurately be used to make inference on process level parameters for self-exciting models and aid analysts in determining the appropriate process model when scientific knowledge cannot be directly applied as in \cite{cressie2015statistics}.  Finally, we will apply this technique to terrorism data in Iraq.  We will show that choice of process model, in this case, results in differing conclusions on the impact of self-excitation in the model.

	\section{Self-Exciting Spatio-Temporal Models}
	The use of self-exciting models in both criminal and terrorism modeling has become increasingly popular over the last decade after being originally introduced in \cite{short2008statistical}.  Self-excitement, in a statistical model, directly models copy-cat behavior by letting an observed event increase the intensity (or excites a model) over a specified time or location. Self-exciting models are closely related to Hawkes processes, which are counting processes where the probability of an event occurring is directly related to the number of events that previously occurred. In a self-exciting model, the criminal intensity at a given spatio-temporal location, $(x,y,t)$ is a mixture of a background rate, $\nu$ and self excitement function, $f(\mathcal{H}_{x,y,t})$ that is dependent on the observed history at that location, $\mathcal{H}_{x,y,t}$.  
	
	A common temporal version of a discretized Hawkes process is
	\begin{align}
		& Y_t\sim \mbox{Pois}(\lambda_t) \\
		& \lambda_t = \nu + \sum_{j < t} \kappa (t-j) y_j \nonumber\\
		& t \in \{1,2,...T\}
	\end{align}

	In this example, in order for the process to have finite expectation in the limit, $\kappa(t-j)$ must be positive and $\sum_{i=1}^{\infty} \kappa (i) <1$.  $\kappa (t-j)$ can be thought of as the probability that an event at time $j$ triggers an event at time $i$.
	
	\cite{laub2015hawkes} provides an excellent overview of the mathematical properties of the continuous Hawkes process and the discrete process when $\kappa (t-j)$ is taken to be an exponential decay function.
	
	In \cite{mohler2012self}, $\nu$ was treated as separable in space and time and was non-parametrically estimated using stochastic declustering, while in \cite{mohler2013modeling}, the spatial correlations were ignored and an AR(1) process was used for $\kappa(t)$ and an exponential decay was assumed for the self-excitation.  In terrorism modeling \cite{lewis2012self} used a piece-wise linear function for $\kappa(t)$.  
	
	Here we will first define a general model that allows spatial or spatial-temporal correlation to exist in the process model and positive temporal correlation to exist in the data model to allow for self-excitation.  First define $s_i \in \mathcal{R}^2$, $i \in \{1,2,...,s\}$ as locations in a fixed, aerial, region. We further define $t \in \{1,2,...,n\}$ as discrete time. The general form of a spatial-temporal self-exciting model is then given in \eqref{eq:gen Model}
	
	\begin{align}
	& Y(\boldsymbol{s_i},t)|\mu(\boldsymbol{s_i},t) \sim \mbox{Pois}(\mu(\boldsymbol{s_i},t)) \label{eq:gen Model}\\
	& \mu(\boldsymbol{s_i},t) = \exp(X(\boldsymbol{s_i},t)) + \eta Y(\boldsymbol{s_i},t-1) \nonumber \\
	& \boldsymbol{X} \sim \mbox{Gau}(\boldsymbol{0},Q^{-1}(\theta))) \nonumber
	\end{align}
	
	Comparing the above to the Hawkes process, we now have $\nu$ as a function of space-time and denote it as $X(\boldsymbol{s_i},t)$.  We use the simplest form of self-excitation letting $\kappa(t-j)$ be a point-mass function such that $\kappa(k)=\eta$ for $k=1$ and $\kappa(k)=0$ for $k \neq 1$.  In all cases $Y(\boldsymbol{s_i},t)$ will be discrete, observable, count data.  
	
\begin{figure}[h]
			\begin{center} 
				\vspace{6pc}
				\includegraphics[width=.5 \linewidth]{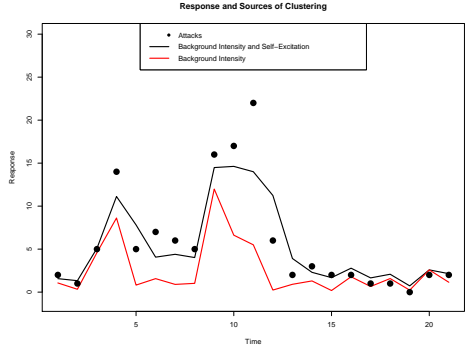}
				\caption[]{This figure shows an example of the expectation of two processes, one with self-excitation and one without.  The bottom line is the expectation of a process withs no self-excitation, the top has self-excitation of $\eta=.4$. The data realizations are from the process with self-excitation.}
				\label{exc} 
			\end{center}
		\end{figure}

	To contrast \eqref{eq:gen Model} with a typical spatial model, figure \ref{exc} depicts the expectation for one aerial location $(\boldsymbol{s_i},t)$ without self-excitation and with self-excitation as shown in Figure \ref{exc}. In this figure, the lower line shows $\mu(\boldsymbol{s_i},t)$ with $\eta=0$, and the upper line has $\eta=.4$.  The impact of self-excitation is clearly present in time 10-13.
	
	\subsection{Spatially Correlated Self-Exciting Model}
	
	In the first example of \eqref{eq:gen Model} we assume the background intensity rate, $X(\boldsymbol{s_i},t)$ has only spatial correlation.  This model is motivated through the assumption that the latent dependency, $X(\boldsymbol{s_i},t)$, is as a continuous measure of violent tendency at region $s_i$ at time period $t$ and regions that are closer together in space are assumed to share common characteristics.  
	
	Next, define $N(\boldsymbol{s_i})$ as the neighborhood of location $s_i$ where two regions are assumed to be neighbors if they share a common border.   $|N(\boldsymbol{s_i})|$ is the number of neighbors of location $s_i$.  The model for $Y(\boldsymbol{s_i},t)$, or the number of observed violent events at a given space-time location is then given by:

	\begin{align}
		& Y(\boldsymbol{s_i},t)|\mu(\boldsymbol{s_i},t) \sim \mbox{Pois}(\mu(\boldsymbol{s_i},t)) \label{eq:Full Model}\\
		& \mu(\boldsymbol{s_i},t) = \exp(X(\boldsymbol{s_i},t)) + \eta Y(\boldsymbol{s_i},t-1) \nonumber \\
		& X(\boldsymbol{s_i},t) = \theta_1 \sum_{\boldsymbol{s_j}\in N(\boldsymbol{s_i})}X(\boldsymbol{s_j},t) + \epsilon(\boldsymbol{s_i},t) \nonumber\\
		&\epsilon(\boldsymbol{s_i},t) \sim \mbox{Gau}(0,\sigma^2) \nonumber
	\end{align}
	
	Letting $\boldsymbol{H}$ denote the spatial neighborhood matrix such that $H_{i,j}=H_{j,i}=1$ if $\boldsymbol{s_i}$ and $\boldsymbol{s_j}$ are neighbors, the full distribution of the joint distribution of the latent state is $\boldsymbol{X}\sim \mbox{Gau} (\boldsymbol{0},(\boldsymbol{I}_{ns,ns}-\boldsymbol{I}_{n,n} \otimes \theta_1\boldsymbol{H})^{-1}\boldsymbol{L}(\boldsymbol{I}_{ns,ns}-\boldsymbol{I}_{n,n} \otimes \theta_1\boldsymbol{H})^{-1})$ where $\boldsymbol{L}=\text{diag}(\sigma^2,...,\sigma^2)$.  The evolution in the latent field is equivalent to the spatial evolution in what is commonly referred to as a Simultaneous or Spatial Auto-regressive model (SAR).  Alternatively, the Conditional Auto-regressive model (CAR) of \cite{besag1974spatial} could be used to model the latent state modifying the joint density above.

	The difference between the SAR and \eqref{eq:Full Model} is in the self excitement parameter, $\eta$.  In \eqref{eq:Full Model}, temporal correlation is present, but is present through the data model specification rather than through a temporal evolution in the latent state.  Therefore, temporal correlation is a function solely of the self-excitation in the process.  $\eta$ gives the probability of an event at time $t-1$ creating an event at time $t$.  In order for the system to be well-behaved, $\eta$ is constrained to (0,1).  In order for the joint distribution of $\boldsymbol{X}$ to be valid, $\theta_1 \in (\psi_{(1)}^{-1},\psi_{(n)}^{-1})$ where $\psi_{i}$ is the $i$th smallest eigenvalues of $\boldsymbol{H}$.
	
	The critical assumption in this model is that the propensity of a given location to be violent is spatially correlated with its adjacent spatial neighbors and only evolves over time as a function of excitation.  If terrorism is diffusing according to this model, regions that are geographically adjacent are behaving in a similar manner.  The existence of self-excitation would indicate that individuals within a region are being inspired through the actions of others.  While combating terrorism is complex, if terrorism is diffusing in this manner, one suggestion would be to identify the root causes within a geographic area as well as quick action against any malicious actor to discourage copy-cat behavior.

	\subsection{Reaction Diffusion Self-Exciting Model}
	
	Alternatively, a model similar to \cite{short2008statistical} can be used to motivate the process model for the latent state resulting in a non-separable spatio-temporal, $\boldsymbol{X}$.  Here we let $X(\boldsymbol{s_i,t})$ corresponds to a continuous measure of violence due to terrorists or criminals at location $\boldsymbol{s_i}$ at time $t$. This is still a latent variable as we are not directly measuring $X(\boldsymbol{s_i,t})$.  However, now in order for an area to increase in violent tendency, a neighboring area must decrease as the actors causing the violence move from location to location.  Furthermore, if terrorists are removed from the battlefield at a rate proportional to the total number of terrorists present and if terrorists move to fill power vacuums, the process model is similar to the reaction-diffusion partial differential equation (see \cite{cressie2015statistics} for more on the reaction-diffusion model)
	\begin{equation}
		\frac{\partial X(\boldsymbol{s_i},t)}{\partial t}=\frac{\kappa}{|N(\boldsymbol{s_i})|} \triangle X(\boldsymbol{s_i},t)-\alpha X(\boldsymbol{s_i},t) \label{eq:Reaction}
	\end{equation}
	 In order to generalize this partial differential equation (PDE) to an irregular lattice, we make use of the graphical Laplacian, $\Gamma$, in place of $\triangle$ in \eqref{eq:Reaction}. $\Gamma$ is a matrix that extends the notion of second derivatives to irregular graphs and can be defined as a matrix of the same dimension as the number of geographical regions with entries given by
	
	\[
	\Gamma (s_i,s_j)
	\begin{cases} 
	\hfill -N(s_i)   \hfill &  j=i \\
	\hfill 1 \hfill & j\in N(s_i)  \\
	\hfill 0 \hfill& \text{Otherwise}
	\end{cases}
	\]
	With the addition of a random noise term assumed to be Gaussian, the full model can be seen as an example of \eqref{eq:gen Model}.
		\begin{align}
		& Y(\boldsymbol{s_i},t)|\mu(\boldsymbol{s_i},t) \sim \mbox{Pois}(\mu(\boldsymbol{s_i},t)) \label{eq:ReacDiffuse Model}\\
		& \mu(\boldsymbol{s_i},t) = \exp(X(\boldsymbol{s_i},t)) + \eta Y(\boldsymbol{s_i},t-1) \nonumber \\
		& \small X(\boldsymbol{s_i},t) = \frac{\kappa}{|N(s_i)|} \sum_{\boldsymbol{s_j}\in N(\boldsymbol{s_i})}X(\boldsymbol{s_j},t-1) + (1-\kappa-\alpha)  X(\boldsymbol{s_i},t-1) + \epsilon(\boldsymbol{s},t) \nonumber\\
		&\epsilon(\boldsymbol{s},t) \sim \mbox{Gau}(0,\sigma^2) \nonumber
		\end{align}
	
	In contrast to the Spatially Correlated Self-Exciting (SCSE) Model, the process model dependency exists in both space and time.  In order to derive properties of this model we first let $\boldsymbol{M}=\kappa \text{ diag}\left( \frac{1}{|N_{s_i}|}\right) \Gamma + (1-\alpha) \boldsymbol{I}_{s,s}$ and now note that this is equivalent to a Vector Auto-Regressive, VAR, model $\boldsymbol{X}_t = \boldsymbol{M} \boldsymbol{X}_{t-1} +  \boldsymbol{\epsilon}$ with $\boldsymbol{\epsilon}\sim \mbox{Gau}(\boldsymbol{0},\sigma^2 \boldsymbol{I})$.
	
	The VAR(1) model requires all the eigenvalues of $\boldsymbol{M}$ to be between -1 and 1.  This can be satisfied by first noting that 0 is always an eigenvalue of $\text{ diag}\left( \frac{1}{|N_{s_i}|}\right)$ trivially corresponding to the eigenvector of all 1s.  The largest eigenvalue is at most 2 as shown in \cite{chung1997spectral}.  Due to the structure of $(1-\alpha) \boldsymbol{I}_{s,s}$ this implies maximum eigenvalue of $\boldsymbol{M}$ is $(1-\alpha)$ and minimum is  $-2\beta+(1-\alpha)$.  Therefore, the parameter spaces for $\alpha$ and $\kappa$ are $\alpha \in (0,1)$ and $\kappa \in (\frac{-\alpha}{2},\frac{2-\alpha}{2})$.

	Just as in the SCSE Model, if $\epsilon$ has a Gaussian distribution, the Reaction Diffusion Self-Exciting (RDSE) Model has an exact solution for the latent Gaussian field, $\boldsymbol{X}$.  
	 
	Letting $\Sigma_s$ be the spatial covariance at a fixed period of time which is assumed to be invariant to time , then we can solve for $\Sigma_s$ by using the relationship $\Sigma_s = \boldsymbol{M}\Sigma_s\boldsymbol{M^T} + \sigma^2 \boldsymbol{I}$.  As demonstrated by \cite{cressie2015statistics}, this leads to  $\text{vec}(\Sigma_s)=\left(\boldsymbol{I}_{s^2,s^2}-\boldsymbol{M}\otimes\boldsymbol{M}\right)^{-1}\text{vec}\left(\sigma^2 \boldsymbol{I}_{s,s}\right)$ where $\text{vec}\left( \right)$ is the matrix operator that stacks each column of the matrix on top of one or another.  Recall that $s$ is the size of the lattice that is observed at each time period.  The joint distribution for all $\boldsymbol{X}$ is then $\boldsymbol{X}\sim \mbox{Gau}(\boldsymbol{0},Q_{rd}^{-1}(\theta))$ where

	\begin{equation}
	Q_{rd}^{-1}(\theta)=
	\left[
	\begin{array}{c|c|c|c}
	\Sigma_s & M \Sigma_s & ... & M^n\Sigma_s \\
	\hline
	\Sigma_s M^T & \Sigma_s & ... & M^{n-1}\Sigma_s\\
	\hline
	... & ... & ... & ...\\
	\hline
	\Sigma_s (M^T)^n & \Sigma_s (M^T)^{n-1}& ... & \Sigma_s
	\end{array} 
	\right]\label{eq:Sig}
	\end{equation}
.

	However, practically, this involves inverting a potentially large matrix $\boldsymbol{I}_{s^2,s^2}-\boldsymbol{M}\otimes\boldsymbol{M}$.  Therefore, it is easier to deal with the inverse of \eqref{eq:Sig} given in \eqref{eq:Prec}.
		\begin{equation}
		Q_{rd}(\theta)=
		\left[
		\begin{array}{c|c|c|c|c}
		\boldsymbol{I}_{n,n} & -M & \boldsymbol{0}&  ... & ... \\
		\hline
		- M^T  & M^T M +\boldsymbol{I}_{n,n} & - M & \boldsymbol{0} & ... \\
		\hline
		\boldsymbol{0} &- M^T  & M^T M +\boldsymbol{I}_{n,n} & - M & ...\\
		\hline
		...&...&...&...&...\\
		\hline
		\boldsymbol{0} & ... & - M^T  & M^T M +\boldsymbol{I}_{n,n} & - M\\
		\hline
		\boldsymbol{0} & ... & ... & -M^T & \boldsymbol{I}_{n,n}
		\end{array} 
		\right] \frac{1}{\sigma^2}\label{eq:Prec}
		\end{equation}

	The primary difference between the SCSE model and the RDSE model is that the process model correlation in the SCSE is only spatial while in the RDSE it is spatio-temporal.  In the below toy examples, we show the expectation for $X(s_i,t)$ for both the SCSE and the RDSE model on a 4 x 4 lattice structure.  We fixed both models with a value of $X(s_1,1)=10$ as the upper left hand observation at time 1.  As seen in the RDSE model, the high count at time 1 spreads to neighboring regions in time 2 and time 3 whereas the process model has no temporal spread in the SCSE but has a high level of spatial spread.
	\begin{center}
				Spatially Correlated Latent Process Conditional on $(s_1,1)=10$
		\begin{tikzpicture}[>=latex']
		\tikzset{block/.style= { rectangle, align=left,minimum width=.2cm,minimum height=.1cm},
			rblock/.style={draw, shape=rectangle,rounded corners=1.5em,align=center,minimum width=.2cm,minimum height=.1cm},
			input/.style={ 
				draw,
				trapezium,
				trapezium left angle=60,
				trapezium right angle=120,
				minimum width=2cm,
				align=center,
				minimum height=1cm
			},
		}
		\node [block, fill=orange!90]  (x1) {\footnotesize 10};
		\node [block, right = .1cm of x1, fill=blue!50] (x2) {\footnotesize 5};
		\node [block, right = .1cm of x2,fill=blue!20] (x3) {\footnotesize 2};
		\node [block, right = .1cm of x3, fill=blue!10] (x4) {\footnotesize 1};
		\node [block, below = .1cm of x1, fill=blue!50] (y1) {\footnotesize 5};
		\node [block, below = .1cm of x2, fill=blue!50] (y2) {\footnotesize 4};
		\node [block, below = .1cm of x3, fill=blue!30] (y3) {\footnotesize 3};
		\node [block, below = .1cm of x4,fill=blue!20] (y4) {\footnotesize 2};
		\node [block, below = .1cm of y1,fill=blue!20] (z1) {\footnotesize 2};
		\node [block, below = .1cm of y2, fill=blue!30] (z2) {\footnotesize 3};
		\node [block, below = .1cm of y3, fill=blue!30] (z3) {\footnotesize 2};
		\node [block, below = .1cm of y4, fill=blue!10] (z4) {\footnotesize 1};
		\node [block, below = .1cm of z1, fill=blue!10] (q1) {\footnotesize 1};
		\node [block, below = .1cm of z2, fill=blue!10] (q2) {\footnotesize 1};
		\node [block, below = .1cm of z3, fill=blue!10] (q3) {\footnotesize 1};
		\node [block, below = .1cm of z4, fill=blue!10] (q4) {\footnotesize 1};
		\node [block, above = .1cm of x2] {\footnotesize Time 1};
		\end{tikzpicture}
		\quad
		\begin{tikzpicture}[>=latex']
		\tikzset{block/.style= { rectangle, align=left,minimum width=.2cm,minimum height=.1cm},
			rblock/.style={draw, shape=rectangle,rounded corners=1.5em,align=center,minimum width=.2cm,minimum height=.1cm},
			input/.style={ 
				draw,
				trapezium,
				trapezium left angle=60,
				trapezium right angle=120,
				minimum width=2cm,
				align=center,
				minimum height=1cm
			},
		}
		\node [block]  (x1) {\footnotesize 0};
		\node [block, right = .1cm of x1] (x2) {\footnotesize 0};
		\node [block, right = .1cm of x2] (x3) {\footnotesize 0};
		\node [block, right = .1cm of x3] (x4) {\footnotesize 0};
		\node [block, below = .1cm of x1] (y1) {\footnotesize 0};
		\node [block, below = .1cm of x2] (y2) {\footnotesize 0};
		\node [block, below = .1cm of x3] (y3) {\footnotesize 0};
		\node [block, below = .1cm of x4] (y4) {\footnotesize 0};
		\node [block, below = .1cm of y1] (z1) {\footnotesize 0};
		\node [block, below = .1cm of y2] (z2) {\footnotesize 0};
		\node [block, below = .1cm of y3] (z3) {\footnotesize 0};
		\node [block, below = .1cm of y4] (z4) {\footnotesize 0};
		\node [block, below = .1cm of z1] (q1) {\footnotesize 0};
		\node [block, below = .1cm of z2] (q2) {\footnotesize 0};
		\node [block, below = .1cm of z3] (q3) {\footnotesize 0};
		\node [block, below = .1cm of z4] (q4) {\footnotesize 0};
		\node [block, above = .1cm of x2] {\footnotesize Time 2};
		\end{tikzpicture}
	\end{center}
	\begin{center}
		Reaction Diffusion Latent Process Conditional on $(s_1,1)=10$
	\end{center}
	\quad
	\begin{center}
		\begin{tikzpicture}[>=latex']
		\tikzset{block/.style= { rectangle, align=left,minimum width=.2cm,minimum height=.1cm},
			rblock/.style={draw, shape=rectangle,rounded corners=1.5em,align=center,minimum width=.2cm,minimum height=.1cm},
			input/.style={ 
				draw,
				trapezium,
				trapezium left angle=60,
				trapezium right angle=120,
				minimum width=2cm,
				align=center,
				minimum height=1cm
			},
		}
		\node [block, fill=orange!90]  (x1) {\footnotesize 10};
		\node [block, right = .1cm of x1,fill=blue!10] (x2) {\footnotesize 1};
		\node [block, right = .1cm of x2, fill=blue!10] (x3) {\footnotesize 1};
		\node [block, right = .1cm of x3] (x4) {\footnotesize 0};
		\node [block, below = .1cm of x1,fill=blue!10] (y1) {\footnotesize 1};
		\node [block, below = .1cm of x2, fill=blue!10] (y2) {\footnotesize 1};
		\node [block, below = .1cm of x3] (y3) {\footnotesize 0};
		\node [block, below = .1cm of x4] (y4) {\footnotesize 0};
		\node [block, below = .1cm of y1, fill=blue!10] (z1) {\footnotesize 1};
		\node [block, below = .1cm of y2] (z2) {\footnotesize 0};
		\node [block, below = .1cm of y3] (z3) {\footnotesize 0};
		\node [block, below = .1cm of y4] (z4) {\footnotesize 0};
		\node [block, below = .1cm of z1] (q1) {\footnotesize 0};
		\node [block, below = .1cm of z2] (q2) {\footnotesize 0};
		\node [block, below = .1cm of z3] (q3) {\footnotesize 0};
		\node [block, below = .1cm of z4] (q4) {\footnotesize 0};
		\node [block, above = .1cm of x2] {\footnotesize Time 1};
		\end{tikzpicture}
		\qquad
		\begin{tikzpicture}[>=latex']
		\tikzset{block/.style= { rectangle, align=left,minimum width=.2cm,minimum height=.1cm},
			rblock/.style={draw, shape=rectangle,rounded corners=1.5em,align=center,minimum width=.2cm,minimum height=.1cm},
			input/.style={ 
				draw,
				trapezium,
				trapezium left angle=60,
				trapezium right angle=120,
				minimum width=2cm,
				align=center,
				minimum height=1cm
			},
		}
		\node [block,fill=blue!30]  (x1) {\footnotesize 3};
		\node [block, right = .1cm of x1, fill=blue!30] (x2) {\footnotesize 3};
		\node [block, right = .1cm of x2] (x3) {\footnotesize 0};
		\node [block, right = .1cm of x3] (x4) {\footnotesize 0};
		\node [block, below = .1cm of x1, fill=blue!30] (y1) {\footnotesize 3};
		\node [block, below = .1cm of x2, fill=blue!10] (y2) {\footnotesize 1};
		\node [block, below = .1cm of x3] (y3) {\footnotesize 0};
		\node [block, below = .1cm of x4] (y4) {\footnotesize 0};
		\node [block, below = .1cm of y1] (z1) {\footnotesize 0};
		\node [block, below = .1cm of y2] (z2) {\footnotesize 0};
		\node [block, below = .1cm of y3] (z3) {\footnotesize 0};
		\node [block, below = .1cm of y4] (z4) {\footnotesize 0};
		\node [block, below = .1cm of z1] (q1) {\footnotesize 0};
		\node [block, below = .1cm of z2] (q2) {\footnotesize 0};
		\node [block, below = .1cm of z3] (q3) {\footnotesize 0};
		\node [block, below = .1cm of z4] (q4) {\footnotesize 0};
		\node [block, above = .1cm of x2] {\footnotesize Time 2};
		\end{tikzpicture}
		\qquad
		\begin{tikzpicture}[>=latex']
		\tikzset{block/.style= { rectangle, align=center,minimum width=.2cm,minimum height=.1cm},
			rblock/.style={draw, shape=rectangle,rounded corners=1.5em,align=center,minimum width=.2cm,minimum height=.1cm},
			input/.style={ 
				draw,
				trapezium,
				trapezium left angle=60,
				trapezium right angle=120,
				minimum width=2cm,
				align=center,
				minimum height=1cm
			},
		}
		
		\node [block, fill=blue!20]  (x1) {\footnotesize 2};
		\node [block, right = .1cm of x1, fill=blue!10] (x2) {\footnotesize 1};
		\node [block, right = .1cm of x2, fill=blue!10] (x3) {\footnotesize 1};
		\node [block, right = .1cm of x3] (x4) {\footnotesize 0};
		\node [block, below = .1cm of x1, fill=blue!10] (y1) {\footnotesize 1};
		\node [block, below = .1cm of x2, fill=blue!10] (y2) {\footnotesize 1};
		\node [block, below = .1cm of x3] (y3) {\footnotesize 0};
		\node [block, below = .1cm of x4] (y4) {\footnotesize 0};
		\node [block, below = .1cm of y1, fill=blue!10] (z1) {\footnotesize 1};
		\node [block, below = .1cm of y2] (z2) {\footnotesize 0};
		\node [block, below = .1cm of y3] (z3) {\footnotesize 0};
		\node [block, below = .1cm of y4] (z4) {\footnotesize 0};
		\node [block, below = .1cm of z1] (q1) {\footnotesize 0};
		\node [block, below = .1cm of z2] (q2) {\footnotesize 0};
		\node [block, below = .1cm of z3] (q3) {\footnotesize 0};
		\node [block, below = .1cm of z4] (q4) {\footnotesize 0};
		\node [block, above = .1cm of x2] {\footnotesize Time 3};
		\end{tikzpicture}\\
	\end{center}

	Practically, if data follows the RDSE model, it implies a high terrorism count in one region will manifest into a high terrorism count in a neighboring region at a later time period.  In combating terrorism, the RDSE might suggest isolating geographical regions to mitigate the risk of spread while addressing self-excitation through direct action against malicious actors who are inspiring others.
	
	\section{Model Fitting}
	In both the RDSE and the SCSE, spatio-temporal diffusion exists in both the process model and the data model.  If the diffusion was solely in the process model, a technique for inference would be Integrated Nested Laplace Approximation, or INLA.
	
	INLA was first proposed in \cite{rue2009approximate} to specifically address the issue of Bayesian Inference of high dimensional Latent Gaussian Random Fields, LGRFs.  An example of this for count data is:
	\begin{align}
		Y(s_i)&\sim \mbox{Pois}(\mu(s_i,t)) \label{eq:model}\\
		\mu(\boldsymbol{s_i,t})&=\exp(\lambda(\boldsymbol{s_i},t))\nonumber\\
		\lambda(\boldsymbol{s_i},t) &= \beta_0 + \boldsymbol{Z}^t \boldsymbol{\beta} + X(\boldsymbol{s_i,t})\nonumber\\
		X(\boldsymbol{s_i,t})& \sim \mbox{Gau}(\boldsymbol{0},Q^{-1}(\theta))\nonumber	
	\end{align}
	
	INLA is often preferable over MCMC for these types of models.  An issue with traditional Markov Chain Monte Carlo (MCMC) techniques for these models is that the dimension of $X$ is often very large.  Therefore, while MCMC has $O_p(N^{-1/2})$ errors, the $N$ in the errors is the simulated sample size for the posterior.  Just getting $N=1$ may be extremely difficult due to the vast number of elements of $X$ that need to be estimated.  In general, MCMC will take hours or days in order to successfully simulate from the posterior making the computational cost of fitting multiple process models extremely high.  In \cite{python2016bayesian}, terrorism data was fit using a grid over the entire planet using INLA, though without self-excitation in the model.
		
	To address the issues with MCMC use in LGRFs, \cite{rue2009approximate} developed a deterministic approach based on multiple Laplacian approximations. A LGRF is any density that can be expressed as
	\begin{equation}
		\pi(\boldsymbol{\theta},\boldsymbol{X} |\boldsymbol{Y}) \propto \pi (\theta)|Q(\boldsymbol{\theta})|^{1/2}\exp \left[\frac{-1}{2}\boldsymbol{X}^t Q(\boldsymbol{\theta}) \boldsymbol{X}+\sum_{\boldsymbol{s}} \log\left(\pi(Y(\boldsymbol{s_i})|X(\boldsymbol{s_i}),\boldsymbol{\theta})\right)\right]
	\end{equation} 
	In order to conduct inference on this model, we need to estimate $\pi(\boldsymbol{\theta}|\boldsymbol{y})$, $\pi(\theta_i|\boldsymbol{y})$ and $\pi(x_i|\boldsymbol{y})$.  The main tool \cite{rue2009approximate} employ is given in their equation (3) as
	\begin{equation}
		\tilde{\pi}(\boldsymbol{\theta}|\boldsymbol{Y})\propto \frac{\pi(\boldsymbol{X},\boldsymbol{\theta},\boldsymbol{Y})}{\tilde{\pi}_G (\boldsymbol{X}|\boldsymbol{\theta},\boldsymbol{Y})}|_{X=x^*(\boldsymbol{\theta})}\label{eq:main}
	\end{equation}
	
	In \cite{rue2009approximate} they note that the denominator of \eqref{eq:main} almost always appears to be unimodal and approximately Gaussian.  The authors then propose to use a Gaussian approximation to $\pi(\boldsymbol{X}|\boldsymbol{\theta},\boldsymbol{Y})$ which is denoted above as $\tilde{\pi}_G$.  Moreover, \eqref{eq:main} should hold no matter what choice of $\boldsymbol{X}$ is used, so a convenient choice for $\boldsymbol{X}$ is the mode for a given $\theta$, which \cite{rue2009approximate} denote as $x^*(\boldsymbol{\theta})$.  
	
	Now, $\pi(\boldsymbol{\theta}|\boldsymbol{Y})$ can be explored by calculating the marginal for choices of $\theta$, which if chosen carefully can greatly decrease the computational time.  These explored values can then be numerically integrated out to get credible intervals for $\pi(\theta_i|\boldsymbol{Y})$.
	
	Following the exploration of $\theta|Y$, and computation of $\theta_i|Y$, INLA next proceeds to approximate $\pi(X(s_i)|\boldsymbol{\theta},\boldsymbol{Y})$.  The easiest way to accomplish this is to use the marginals that can be derived straightforwardly from $\tilde{\pi}_G(\boldsymbol{X}|\boldsymbol{\theta},\boldsymbol{Y})$ from \eqref{eq:main}.  In this manuscript we will use this technique for simplicity of computation, however, if the latent states are of interest in the problem (and they often are), this can be problematic as it fails to capture any skewness of the posterior of $\boldsymbol{X}$.  One way to correct this is to re-apply \eqref{eq:main} in the following manner:
	\begin{equation}
		\tilde{\pi}_{LA}(X(\boldsymbol{s_i})|\theta,y)\propto \frac{\pi(\boldsymbol{X},\theta,\boldsymbol{Y})}{\tilde{\pi}_G(\boldsymbol{X}_{-s_i}|X(s_i),\boldsymbol{\theta},\boldsymbol{Y})}|_{x_{-i}=\boldsymbol{x_{-i}}^*(x_i,\theta)}\label{eq:remain}
	\end{equation}
	In \eqref{eq:remain} $\boldsymbol{X}_{-s_i}$ is used to represent $\boldsymbol{X}$ with latent variable $X(s_i)$ removed.  This is a reapplication of Tierney and Kadane's marginal posterior density and gives rise to the nested term in INLA. 
	
	\subsection{Laplace Approximation for Spatio-Temporal Self-Exciting Models}
	While INLA is an attractive technique due to computational speed and implementation, it is not immediately usable for the SCSE and the RDSE as the structure in \eqref{eq:gen Model} is
		\begin{align}
		\mu(\boldsymbol{s_i},t) & = \exp(X(\boldsymbol{s_i},t)) + \eta Y(\boldsymbol{s_i},t-1)\nonumber\\
		& \eta \in (0,1)
		\end{align}
	In this structure, $X(.)$ and $Y(.)$ are not linearly related and a Gaussian prior for $\eta$ is clearly not appropriate due to the parameter space constraints.

	However, Laplace approximations can still be used by conducting inference on $\eta$ at the same time inference is conducted on the the set of latent model parameters.  In both the Spatially Correlated Self-Exciting Model and the Reaction Diffusion Self-Exciting model, the full conditional for the latent state is
	\begin{equation}
		\footnotesize\pi(\boldsymbol{X}|\boldsymbol{Y},\boldsymbol{\theta}) \propto \exp\left(-\frac{1}{2}\boldsymbol{X}^T \boldsymbol{Q(\boldsymbol{\theta})}\boldsymbol{X} + \sum_{s_i,t} \log \pi\left(Y(\boldsymbol{s_i},t)|X(\boldsymbol{s_i},t),\eta,Y(\boldsymbol{s_i},t-1)\right)\right)\label{eq:FullCond}
	\end{equation} 
	Here we will let $\boldsymbol{\theta}=(\theta_1,\sigma^2,\eta)^T$ and $\boldsymbol{Q_{sc}(\boldsymbol{\theta})}=(\boldsymbol{I}_{sn,sn}-\theta_1\boldsymbol{I}_{t,t} \otimes \boldsymbol{H})$ for the Spatially Correlated Self-Exciting Model and use $\boldsymbol{Q_{rd}(\boldsymbol{\theta})}$ for the RDSEM defined in \eqref{eq:Prec}.
	
	  While $\boldsymbol{\theta}$ in \eqref{eq:FullCond} does not contain $\eta$ we next do a Taylor series expansion of $\log \pi\left(Y(\boldsymbol{s_i},t)|X(\boldsymbol{s_i},t),\eta,Y(\boldsymbol{s_i},t-1)\right)$, as a function of $X(\boldsymbol{s_i},t)$ and, for each $\boldsymbol{s_i},t$, expand the term about a guess for the mode, say $\mu_0(\boldsymbol{s_i,t})$.  First we write $\boldsymbol{B^*}(\boldsymbol{\theta}|\mu_0)$ as a vector of the same length as $X(\boldsymbol{s_i},t)$ where each element is given by
	\begin{equation}
	B(\boldsymbol{s_i},t|\mu_0)=\left(\frac{\partial \log\pi\left(Y(\boldsymbol{s_i},t)\right)}{\partial X(\boldsymbol{s_i},t)}\Bigr|_{X(\boldsymbol{s_i},t)=\mu(\boldsymbol{s_i},t)}-\mu(\boldsymbol{s_i},t) \frac{\partial^2\log\pi\left(Y(\boldsymbol{s_i},t)\right)}{\partial X(\boldsymbol{s_i},t)^2}\Bigr|_{X(\boldsymbol{s_i},t)=\mu(\boldsymbol{s_i},t)}\right)\label{eq:B(si)}
	\end{equation}
	
	Next, we further define
	$\boldsymbol{Q^*(\boldsymbol{\theta})}|\mu_0$ as the updated precision matrix.
	\begin{equation}
	\boldsymbol{Q^*(\boldsymbol{\theta})|\mu_0}=\boldsymbol{Q(\boldsymbol{\theta})}+\text{diag }\left(-\frac{\partial^2\log \pi\left(Y(\boldsymbol{s_i},t)\right)}{\partial X(\boldsymbol{s_i},t)^2}\right)\Bigr|_{X(\boldsymbol{s_i},t)=\mu(\boldsymbol{s_i},t)} \label{eq:Precision at Mode}\\
	\end{equation}
	Where $\boldsymbol{Q(\boldsymbol{\theta})}$ is either $\boldsymbol{Q_{sc}(\boldsymbol{\theta})}$ or $\boldsymbol{Q_{rd}(\boldsymbol{\theta})}$ depending on the context.  Then we can write
	\begin{equation}
		\footnotesize\pi(\boldsymbol{X}|\boldsymbol{Y},\boldsymbol{\theta}) \propto \exp\left(-\frac{1}{2}\boldsymbol{X}^T\left( \boldsymbol{Q^*(\boldsymbol{\theta})|\mu_0}\right)\boldsymbol{X} + \boldsymbol{X}^T\left( \boldsymbol{B^*}(\boldsymbol{\theta})|\boldsymbol{\mu_0}\right)\right) \label{eq:FullCondExpand}
	\end{equation}

	While in \eqref{eq:Precision at Mode}, $\boldsymbol{Q}(\boldsymbol{\theta})$, the original precision matrix, does not contain $\eta$, $\boldsymbol{Q^*(\boldsymbol{\theta})}$, the updated precision matrix, does depend on the self-excitation parameter.

	Next we find the values of $\mu(\boldsymbol{s_i})$ that maximize \eqref{eq:FullCondExpand}.  This is done through the use of an iterative maximization algorithm by solving for $\boldsymbol{\mu_1}$ in $\left(Q^*(\boldsymbol{\theta})|\boldsymbol{\mu_0}\right)\boldsymbol{\mu_1}=\boldsymbol{B^*}(\boldsymbol{\theta}|\mu_0)$. For a fixed $\boldsymbol{\theta}$, this converges rapidly, due to the sparsity of both $\boldsymbol{Q_{sc}}$ and $\boldsymbol{Q_{rd}}$.  .
	
	In \eqref{eq:main}, for a fixed $\boldsymbol{\theta}$, we can then find $x^*(\boldsymbol{\theta})$.  When the denominator of \eqref{eq:main} is evaluated at $x^*(\boldsymbol{\theta})$ it becomes $|\boldsymbol{Q^*(\boldsymbol{\theta})}\frac{1}{2\pi}|^{1/2}$ which is equivalent to the hyperparameter inference recommended by \cite{lee1996hierarchical} as pointed out by R. A. Rigby in \cite{rue2009approximate}.  
	
	In order to best explore $\pi(\boldsymbol{\theta}|\boldsymbol{Y})$ the posterior mode is first found through a Newton-Raphson based method.  In order to do this we approximate the Hessian matrix based off of finite difference approximation to the second derivatives.

	After locating the posterior mode of $\pi(\boldsymbol{\theta}|\boldsymbol{Y})$, the parameter space can be explored using the exploration strategy laid out in section 3.1 of \cite{rue2009approximate}.
	
	Now, for the set of diffusion parameters, $\boldsymbol{\theta}$ which contain $\eta$, we have a method of estimating $\pi(\boldsymbol{\theta}|\boldsymbol{Y})$.  Inference for any further data model covariates can now be conducted in the same manner as done in \cite{rue2009approximate}. 
	\subsection{Model Comparison and Goodness of Fit}
	In order to conduct model comparison, we will use the deviance information criterion (DIC) originally proposed by \cite{spiegelhalter2002bayesian}.  Goodness of fit will be conducted through the use of posterior predictive p-values, outlined by \cite{gelman1996posterior}.
	
	To approximate the DIC, we first find the effective number of parameters for a given $\boldsymbol{\theta}$.  As noted in \cite{rue2009approximate}, we can estimate this by using $n-\text{tr}\left(\boldsymbol{Q(\theta)}\boldsymbol{Q^*(\theta)}^{-1}\right)$ for both the SCSEM and the RDSEM.  This gives the effective number parameters for a given $\boldsymbol{\theta}$, which can then be averaged over $\pi(\boldsymbol{\theta}|\boldsymbol{Y})$ to get the effective number of parameters for the model.
	
	Secondly, we calculate the deviance of the mean
	\begin{equation}
		-2\sum_{s_i,t} \log \pi\left(Y(s_i,t)|\hat{X}(s_i,t),\boldsymbol{\theta^*}\right)
	\end{equation} where $\boldsymbol{\theta^*}$ is the posterior mode and $\hat{X}(s_i,t)$ is the expectation of the latent state fixing $\theta=\theta^*$.  DIC can then be found through deviance of the mean plus two times the effective number of parameters as in chapter 7 of \cite{gelman2014bayesian} and initially recommended by \cite{spiegelhalter2002bayesian}.
	
	In order to assess goodness of fit in analyzing the terrorism data in Section 5, we will use posterior predictive P-values as described by \cite{gelman1996posterior}.  Here, we pick critical components of the original dataset that we wish to see if the fitted model can accurately replicate, for instance the number of zeros in the dataset which we can designate as $T(\boldsymbol{Y})$.  Next, for an index $m=1...M$, We then draw a value of $\boldsymbol{\theta_m}$ according to $\pi(\boldsymbol{\theta}|\boldsymbol{Y})$ and simulate a set of observations $Y^*(\boldsymbol{s_i,t})_m$ of the same dimension as $\boldsymbol{Y}$ and compute $T(\boldsymbol{Y}^*_m)$.  This process is repeated M times and a posterior predictive p-value is computed as $\frac{1}{M}\sum_{m=1}^M I\left[T(\boldsymbol{Y}^*_m) > T(\boldsymbol{Y}) \right]$ where $I\left[ . \right]$ is the indicator function.  While not a true P-value, both high and low values of the posterior predictive p-value should cause concern over the fitted models ability to replicate features of the original dataset.
	
	\section{Simulation}

	In order to validate the Laplace based methodology for spatially correlated self-exciting models we conducted simulation studies using data on a 8 by 8 Spatial grid assuming a rook neighborhood structure.  In order to decrease the edge effect, we wrapped the grid on a torus so each node had four neighbors.  For each grid location we simulated 100 observations, creating a spatio-temporal model that had 6400 observations, meaning in \eqref{eq:main}, $\boldsymbol{Q}(\boldsymbol{\theta})$ had a dimension of $6400 \times 6400$.
	
	In the first simulation we used \eqref{eq:Full Model} fixing the parameters at values that generated data that appeared to resemble the data from Iraq used in Section 5.  The generating model we used was:
	
	\begin{align}
	& Y(\boldsymbol{s_i},t)|\mu(\boldsymbol{s_i},t) \sim \text{Pois }(\mu(\boldsymbol{s_i},t)) \label{eq:First Simulation}\\
	& \mu(\boldsymbol{s_i},t) = \exp(-1+X(\boldsymbol{s_i},t)) + .2 Y(\boldsymbol{s_i},t-1) \nonumber \\
	& X(\boldsymbol{s_i},t) = .22 \sum_{\boldsymbol{s_j}\in N(\boldsymbol{s_i})}X(\boldsymbol{s_j},t) + \epsilon(\boldsymbol{s_i},t) \nonumber\\
	&\epsilon(\boldsymbol{s_i},t) \sim Gau(0,.4) \nonumber
	\end{align}
	
	 The spatial parameter for model was $\theta_1=.22$ which suggests a positive correlation between spatially adjacent locations.  An $\eta$ value of 0.2 would suggest that each event that occurs at one time period increases the expected number of events at the next time period by .2.  Here we fix $\sigma^2$ was fixed at 0.4 and use a value of $\beta_0=-1$  to reflect that in real world applications the latent process likely is not zero mean.
	 
	Once the data were generated, we found $\pi(\boldsymbol{\theta}|\boldsymbol{Y})$ by applying \eqref{eq:main}.  Here we note that the numerator of \eqref{eq:main} is $\small\pi(\boldsymbol{X},\boldsymbol{\theta},\boldsymbol{Y})=\pi(\boldsymbol{Y}|\boldsymbol{X},\eta,\boldsymbol{\theta})\pi(\eta)\pi(\boldsymbol{X}|\theta_1,\sigma^2)\pi(\theta_1)\pi(\sigma^2)$ which requires a prior specification for $\eta$,$\theta_1$, and $\sigma$.  In order to reflect an a-priori lack of knowledge we choose vague priors for all parameters.   In this model, we use a Half-Cauchy with scale parameter of 25 for $\sigma$ and a Uniform ($\psi_{1}^{-1},\psi_{n}^{-1}$) where $\psi_(i)$ is the $i$th largest eigen vector of the spatial neighborhood.  As we used a shared-boreder, or rook, neighbor structure wrapped on a torus, the parameter space is (-0.25,.025) as each spatial location has four neighbors.  The choice of the Half-Cauchy is in line with the recommendations for vague priors for variance components of hierarchical models as outlined in \cite{gelman2006prior} and rigorously defended in \cite{polson2012half}.  We let the prior for $\eta$ be Uniform(0,1).  
	
	Using a gradient descent method with step-halving we found the posterior mode of $\pi(\boldsymbol{\theta}|\boldsymbol{Y})$ to be $\sigma^2=0.32$, $\theta_1=0.22$, and $\eta=0.20$. Using the z based parameterization described in Section 3.1 we next explored the parameterization $\log\pi(\boldsymbol{\theta}|\boldsymbol{Y})$ and found credible intervals of $\pi(\sigma^2|\boldsymbol{Y})=(0.29,0.36),\pi(\theta_1|\boldsymbol{Y})=(0.22,0.23)$ and $\pi(\eta|\boldsymbol{Y})=(.18,.21)$.  Fixing $\boldsymbol{\theta}$ at the posterior mode, we then found an approximate 95\% credible interval for $\beta_0$ to be (.07,-1.67).  In further refinement for $\beta_0$ was required, we could proceed to use \eqref{eq:remain}.  This was not done here as $\beta_0$ was not the subject of our primary inference.
	
	The posterior maximum and credible interval for $\sigma^2$ appear to be slightly lower than expected, but the remaining parameter credible intervals covered the generating parameter.  
	
	Next we simulated from the reaction-diffusion self-excitation model letting $\beta_0=0$, $\alpha=0.1$, $\kappa=0.2$, $\sigma^2=0.25$, and $\eta=0.4$ 
	
	In fitting the model, we again use vague priors for all the parameters.  Again, we place a Half-Cauchy prior on $\sigma^2$ as described above.  In order to conform to the parameter space of $\alpha$ and $\kappa,$ we let $\pi(\alpha)\sim \text{Unif }(0,1)$ and $\pi(\kappa|\alpha)\sim\text{Unif }(-\frac{\alpha}{2},1-\frac{\alpha}{2})$.

	Again using the Laplace approximation technique of section 3, we found the posterior mode of $\pi(\boldsymbol{\theta}|\boldsymbol{Y})$ to be at $\alpha=0.085$, $\kappa=0.19$, $\sigma^2=0.21$, $\eta=0.35$.  95\% credible intervals for the posterior marginals were $\alpha \in (0.07, 0.10)$, $\kappa \in (0.14,0.24)$, $\sigma^2 \in (0.18,0.24)$, and $\eta \in (0.32, 0.40)$.  At the posterior mode of $\boldsymbol{\theta}$, the  posterior marginal for $\beta_0$ was approximately (-0.03,0.01).  Critically, if there is self-excitement in the data, in all simulations it was differentiable from the latent diffusion.  This is a spatial-temporal analogue to the finding in \cite{mohler2013modeling} where a temporal AR(1) process was differentiable from self-excitement.

	In our simulations, the approximations described in this manuscript performed reasonably well for inference on the spatio-temporal diffusion parameters in most cases.  However, when $\sigma^2$ is large, or when $\eta$ is large, we have found that the approximations create bias in one or more of the parameters likely due to the high effective number of parameters.  However, all approximate likelihood based methods will likely struggle in these cases as well.  As noted in \cite{rue2009approximate}, the approximation error in Laplace based methods is related to the number of effective latent variables over the total sample size.

	\section{Spatio-Temporal Diffusion of Violence in Iraq (2003-2010)}
	
	\subsection{Statistical Models and Data}
	
	One region where the reasons for the diffusion of terror and crime still remains unclear is in Iraq during 2003 to 2010.  While violence undoubtedly spread throughout the country, it remains unclear how or why, spatio-temporally, the spread occurred.  Part of the uncertainty is that there still is not agreement over whether violence was due to insurgency, civil war, or organized crime. For example, \cite{hoffman2006insurgency} refers to the violence in Iraq as an insurgency, \cite{fearon2007iraq} argues that the spread of violence was due to a civil war, and \cite{williams2009criminals} argues that there was a large presence of organized crime in the country.  
	
	A few previous studies have examined the presence or absence of self-excitaiton.  In \cite{lewis2012self}, the authors concluded that self-excitation was present in select cities in Iraq during this time period.  The presence of the self-excitation finding was echoed in \cite{braithwaite2015battle} where the authors also noted a correlation between locations that shared microscale infrastructure similarities.  This would suggest repeat or near-repeat actions were causing the increase in violence in a region.
	
	However, in both of these cases, the latent spatio-temporal diffusion was, a-priori, assumed to be known.  In fact, this is likely not the case. In a classic work on the subject, \cite{midlarsky1980violence} discuss how heterogeneity between locations can cause correlation in violence or individuals who cause violence can actually physically move from one location to another.  In particular, if violence is strictly due to crime we would expect self-excitement and limited diffusion between geographical regions.  Whereas if violence is due to insurgencies we would expect more movement of actors as they seek to create widespread disruption in the country.
	The former theory is reflected in the Spatial Correlation in the Spatially Correlated Self-Exciting model and the later theory would correspond to the Reaction Diffusion component of the second model.  
	
	The overarching goal of this analysis, thus, is to determine whether in Iraq the growth of violence in fixed locations was due to the presence or absence of self-excitation.  Furthermore, we want to determine whether the latent diffusion of violence is due to the movement of population such as in the Reaction Diffusion model, or whether there is static spatial correlation.  We will answer this while controlling for exogenous factors that may also explain the rise in violence in a region.

	In order to address this question as well as demonstrate how Laplace based approximations can be used to fit real world data to models of the class of \eqref{eq:model} we used data from the Global Terrorism Database (GTD) introduced in \cite{lafree2007introducing} to examine the competing theories.  The GTD defines terrorist events as events that are intentional, entail violence, and are perpetrated by sub-national actors.  Additionally, the event must be aimed at obtaining a political, social, religions, or economic goal and must be conducted in order to coerce or intimidate a larger actor outside of the victim.  The majority of lethal events in Iraq from 2003-2010 fit the above category.
	
\begin{figure}[h] 
		\begin{center}
			\vspace{6pc}
			\includegraphics[width=.8 \linewidth]{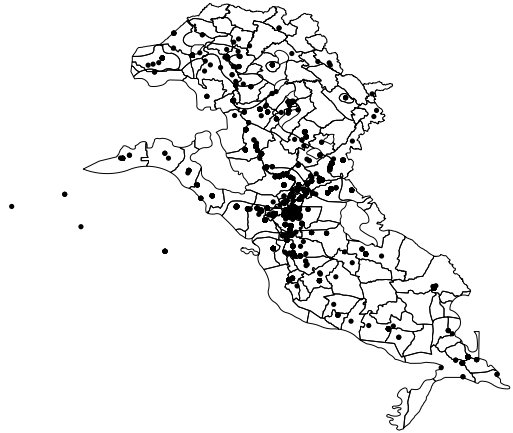}
			\caption[]{Spatial depiction of 6263 events in Iraq.}
			\label{SpaceIZ}
		\end{center}
	\end{figure}

	The GTD uses a variety of open media sources to capture both spatial and temporal data on terroristic events.  The database contains information on what the event was, where it took place, when it took place, and what terrorist group was responsible for the event.  From 2003 to 2010 in Iraq, the database contains 6263 terrorist events, the spatial structure is shown in figure \ref{SpaceIZ}.  
	
	As seen in this map, the majority of the violence is in heavily populated areas such as Baghdad and in the regions north up the Tigris river to Mosul and west through the Euphrates river.  In order to model this data, we aggregated the point data to 155 political districts intersected by ethnicities and aggregated the data monthly for 96 months meaning that $\Sigma(\theta)$ in \eqref{eq:gen Model} is a 14880 x 14880 matrix.  Population for each political district was taken from the Empirical Studies of Conflict Project website, available from https://esoc.princeton.edu/files/ethnicity-study-ethnic-composition-district-level.  
	
	We considered covariates controlling for the population density within a fixed region as well as for the underlying ethnicity.  We will make the simplifying assumption that both of these are static over time.  Previous statistical analysis on terrorism considered macro level covariates, such as democracy in \cite{python2016bayesian} that differ country to country but would not change within a single country as analyzed here.  Other studies considered more micro level covariates such as road networks that were found to be statistically related to terrorism in \cite{braithwaite2015battle}.  Here we take the view point that the vast majority of the incidents in Iraq were directed against individuals rather than terrorist events directed at fixed locations.  Therefore, we would expect a higher population density to provide more targets for a potential terrorist to attack.  Furthermore, covariates such as road networks, number of police, or number of US soldiers, would all be highly collinear with population density.  We do, though, consider a covariate for ethnicity in a region.  Specifically, we add an indicator if the region is predominately Sunni.    The disenfranchisement of the Sunnis and high level of violence in Sunni dominated areas has been well established, see for e.g. \cite{baker2006iraq}.  Previous research in \cite{linke2012space} focusing on Granger Causality also suggested indicators for majority Sunni/Shia may be appropriate in any analysis of violence in Iraq.

	\begin{align}
		& Y(\boldsymbol{s_i},t)|\mu(\boldsymbol{s_i},t) \sim \text{Pois }(\mu(\boldsymbol{s_i},t)) \label{eq:IZ Model}\\
		& \mu(\boldsymbol{s_i},t) = \exp[\beta_0+\beta_1\log \text{Pop}(\boldsymbol{s_i})+\beta_2\text{Sunni }(\boldsymbol{s_i})+X(\boldsymbol{s_i},t)] + \eta Y(\boldsymbol{s_i},t-1) \nonumber \\
		& X(\boldsymbol{s_i},t)\sim \mbox{Gaus}(\boldsymbol{0},\boldsymbol{Q}^{-1}(\theta))  \nonumber
	\end{align}
	
	The complete statistical model is given in \eqref{eq:IZ Model}.  We next fit this model letting $\boldsymbol{Q}=\boldsymbol{Q_{sc}}$ and $\boldsymbol{Q}=\boldsymbol{Q_{rd}}$.  We further consider fixing $\eta=0$ to test the presence or absence of self-excitement in the data for both process models.

	\subsection{Results}

	We fit all four models using the Laplace approximation method described in Section 3.1.  For each of the parameters we used vague proper priors to ensure posterior validity.  In the SCSEM and the Spatially Correlated models we used a Half-Cauchy with scale parameter of 5 for $\sigma$ and a Uniform ($\psi_{1}^{-1},\psi_{n}^{-1}$) where $\psi$ are the eigenvalues of $\boldsymbol{H}$.  Using the neighborhood structure corresponding to the geographical regions described above, this corresponded to a Uniform (-.3,.13). For each of the exogenous covariates, we used independent Gaussian (0,1000) priors.  For the SCSEM model we further assumed a Uniform (0,1) prior for $\eta$.
	
	In fitting the RDSEM, we again used a Half-Cauchy with scale parameter of 5 for $\sigma$.  For the decay parameter, $\alpha$, we used a Uniform (0,1) prior and for the diffusion parameter, $\kappa$, we chose a Uniform $\left(\frac{-\alpha}{2},1-\frac{\alpha}{2}\right)$ in order to ensure we were in the allowable parameter space.

		\begin{table}[h]
			\caption {95\% Credible Intervals for Model Parameters} \label{tab:params} 
			\begin{center}
				\begin{tabular}{ |c|c|c|c|c| } 
					\hline
					 & Spatial Correlation Only & SCSEM & Reation Diffusion Only & RDSEM \\
					\hline 
					$\beta_0$&(-20.2,-18.4) &(-16, -15.5) &(-22.4,-18.0) &(-22,-18.6 )\\
					$\beta_1$& (1.2,1.4)&(0.9, 1.1) &(1.1, 1.4) &(1.1, 1.5)\\
					$\beta_2$& (1.6,1.9)&(1.1,1.3) &(0.10, 0.33) &(0.17, 0.53)\\
					$\eta$& - &(0.35,0.37) & -  &(0, 0.04)\\
					$\sigma^2$&(1.9,2.7) &(2.1, 2.5) & (0.20, 0.30)&(0.18,0.26)\\
					$\theta_1$&(.08,.10) &(0.09,0.1) & - & - \\
					$\alpha$& - & - & (0.001, 0.007) &(0.001, 0.007)\\
					$\kappa$& - & - & (0.03, 0.07) &(0.03, 0.06) \\
					\hline
				\end{tabular}
			\end{center}
			\end{table}

	All four models took approximately 30 min to an hour depending on starting values to converge using a Newton-Raphson based algorithm to find the maximum.  Gaussian approximations to the 95\% credible intervals for the parameters are given for all four models are shown in table \ref{tab:params}.  As can be seen in comparing the SCSEM to the RDSEM, the presence or absence of self-excitation appears to be dependent on the choice of structure of $\boldsymbol{Q}$.  Furthermore, the impact of majority Sunni is also dependent on whether the Reaction Diffusion or Spatially Correlated model was used.
		
	Using the methodology described in Section 3.2, we next calculated DIC as well as posterior predictive P-values based off of the maximum observed value and the number of zeros in the dataset.  In the original dataset, the maximum number of events observed for all regions and months was 26 and the dataset had 13445 month/district observations that were zero. The model assessment and selection results are shown in Table \ref{tab:results}.
	\begin{table}
	\caption {Model Assessment and Selection Statistics For Iraq Data} \label{tab:results} 
	\begin{center}
		\begin{tabular}[H]{|c|c|c|c| } 
			\hline
			Model & DIC & \begin{tabular}{@{}c@{}}P-Values\\ Maximum Value \end{tabular} & \begin{tabular}{@{}c@{}}P-Value \\ Zeros\end{tabular} \\
			
			\hline
			\begin{tabular}{@{}c@{}}Spatially Correlated Model \\ without Self-Excitation \end{tabular} & 9370 & .02 & 1 \\
			\hline	
			\begin{tabular}{@{}c@{}}Spatially Correlated \\ Self-Exciting Model \end{tabular} &  8722 & .97 & 0\\
			\hline 
			\begin{tabular}{@{}c@{}}Reaction Diffusion \\ Only Model \end{tabular} & 8664 & .53 & .81\\
			\hline
			\begin{tabular}{@{}c@{}}Reaction Diffusion \\ Self-Exciting Model \end{tabular} & 8699 & .45 & .89\\
				\hline
		\end{tabular}
	\end{center}
	\end{table}
	
	Clearly from table \ref{tab:results}, the models with an underlying reaction diffusion process model outperform those with spatial correlation only.  Furthermore, the addition of self-excitation in the model appears to have minimal impact.  In particular, without self-excitation, the spatial correlation model tends to under count the number of violent activities while the SCSEM tends to over count.  There really is not much difference between the RDSEM and the reaction diffusion model so we prefer the simper reaction diffusion only model. While $\beta_2$ is only minimally significant in the reaction diffusion model, the models perform better including the covariate than disregarding it entirely.
	
	\subsection{Significance}
	
	Under all measures of performance, the reaction diffusion model, \eqref{eq:ReacDiffuse Model}, without self-excitation outperforms the other models under consideration.  This process model as well as values of the covariates and the lack of self-excitement in the data offer several insights into the causes of the spread of violence in Iraq.
	
	Not surprisingly, the reaction diffusion model has a positive relationship between log population and violence.  As the majority of attacks are directed at individuals it would clearly follow that regions that have higher population will offer more targets as well as more potential combatants.  Further, higher populated areas also would have had higher number of Iraqi government officials as well as US military presence.  The positive, though small, relationship between Sunni and violence is also not surprising as, in general, predominately Sunni areas were generally more disenfranchised following the transition to a new Iraqi government after the downfall of Saddam Hussein.
	
	More significantly, though, was the finding that the reaction diffusion model fit the data better than the SCSEM or the RDSEM.  This suggests that increases in violence can be attributed, at least in part, to movement between high violence and low violence areas rather than repeat or near-repeat actors in a fixed location.

\begin{figure}[h]
			\vspace*{.51cm}
			\centering
			\includegraphics[width=8cm]{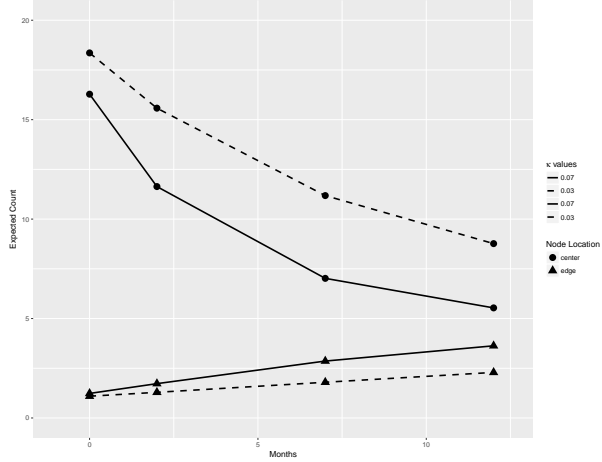}
			\caption{This plot shows the expected changes in violence in a simple three node system where the center node starts with a high level of violence and the other two nodes start with a low level of violence for $\kappa=.03$, depicted as dashed lines in above figure, and $\kappa=.07$, depicted as straight lines.  As seen here after 12 months for $\kappa=.07$ the nodes are essentially at equilibrium. }
			\label{fig:kappa}
		\end{figure}

	While the $\kappa$ parameter may appear small in \ref{tab:params} it still has an impact on the process.  For the sake of simplicity, we can demonstrate this on a three node system.  For this system we consider a central node that has a high level of violence surrounded by two nodes that have a low level of violence. In this set up we will fix $\beta_0=-19$, $\beta_1=1.3$ and consider each node as having a population of 1000 and let $\kappa \in \{.03,.07\}$.  The resultant system over time is shown in Figure \ref{fig:kappa}.   Even in this simple system, there is a noticeable increase in violence as the center node diffuses throughout the entire system.

	The implications of the reaction-diffusion model being preferable over the SCSEM or the spatial correlation only model can be seen by going back to the original PDE that inspired the model, \eqref{eq:Reaction}.  The underlying assumption in that model is that violence is that the rate of violence spreading to a region is determined through the levels of violence in neighboring region.  From a military planning perspective, this would suggest that if there is a peaceful region surrounded by areas of high violence, the peaceful region should be isolated to prevent the movement of malicious actors.  This strategy would be consistent with published military strategy as outlined in \cite{army2006counterinsurgency}.

	Finally, this offers insight into the nature of the conflict that was fought in Iraq.  For instance, \cite{zhukov2012roads} discuss how insurgencies diffuse throughout a population by either physical movement of actors or through movement of ideas, whereas \cite{short2008statistical} suggest that criminal violence would be expected to have an element of self-excitation.  When accounting for the possibility of spatial diffusion, this self-excitation does not appear to be present in the Iraqi dataset. Therefore, as a diffusion based model fits the data better, this would suggest the spread of violence was due to the physical movement of an insurgent population or ideology rather than a criminal element that would be expected to stay more static at a location.

	\section{Discussion}
	
	In this manuscript we develop statistical models that allow for spatio-temporal diffusion in the process model and temporal diffusion in the data model.  We relate the models to existing theory in how violence diffuses in space and time.  We further developed a Laplace approximation for spatio-temporal models that contain self-excitement.  This modification allows for a quick and accurate fit to commonly used models in both the analysis of terrorism and criminology. A critical difference between classical INLA and our proposal is that in our proposal, inference is not only performed on the hyperparameters during the exploration of $\pi(\theta|Y)$ in \eqref{eq:main} but also on the self-excitation parameter $\eta$.  While $\eta$ is not generally thought of as a hyperparameter, when the linear expansion of the log-likelihood is done in \eqref{eq:FullCondExpand}, $\eta$ enters into $\boldsymbol{Q^*}$ and $\boldsymbol{B^*}$ in a similar manner as the hyperparameters. 

	While we only considered two process models and self-excitation that only exists for one time period, the methodology outlined above can easily be extended to allow self-excitation to have an exponential decay similar to the modeling technique of \cite{mohler2012self}.  As shown above, the absence of testing multiple process models may result in premature conclusions about how violence is spreading over regions.  While self-excitation may be present in one model, its significance may be dulled through the use of alternative process models resulting in differing conclusions.
	
	Although self-excitation has become increasingly popular, alternate approaches based on Besag's auto-logistic model, as used in \cite{weidmann2010predicting} are possible, though care must be taken if count data is used as Besag's auto-Poisson does not permit positive dependency. As shown in \cite{kaiser1997modeling}, a Winsorized Poisson must be used if positive dependency is desired, as it most certainly is in terrorism modeling.  In this case, the data model dependency would linearly be associated with the log of $\mu(\boldsymbol{s_i},t)$.
	
	Though the motivation for the models in this manuscript was the spatio-temporal spread of violence, the novel concept of combining latent process dependency and data model dependency has the potential to be used in other fields.  For example in the modeling of thunderstorms, self-excitation may be present temporally, while process model dependency may also be appropriate due to small-scale, unobservable, spatial or spatio-temporal dependency.  Laplace approximations, as demonstrated in this manuscript, allow for quick and relatively accurate methods to fit multiple types of self-exciting spatio-temporal models for initial inference.

	\bibliography{BibFileIz}
\end{document}